\documentstyle[11pt,fleqn,epsf]{article}
\def\ref{par\noindent\hangindent=6mm\hangafter=1}
\begin{document}
\vbox{
%\rightline{IFUG/9602RS1}
\rightline{gr-qc/9603058}
}
\baselineskip 8mm
%\draft

\begin{center}
{\bf One-parameter family of closed, radiation-filled Friedmann-Robertson-Walker 
``quantum" universes }

\bigskip

H.C. Rosu\footnote{Electronic mail:
rosu@ifug.ugto.mx}$^{\dagger}$$^{\ddagger}$
%\cite{byline}
and J. Socorro\footnote{Electronic mail:
socorro@ifug.ugto.mx}$^{\dagger}$
%\cite{byline}

$^{\dagger}$
{\it Instituto de F\'{\i}sica de la Universidad de Guanajuato, Apdo Postal
E-143, L\'eon, Gto, M\'exico}

$^{\ddagger}$
{\it Institute of Gravitation and Space Sciences, P.O. Box MG-6,
Magurele-Bucharest, Romania}

\end{center}

\bigskip
\bigskip

\begin{abstract}

Using as an illustrative example the $p$=1
operator-ordered Wheeler-DeWitt equation for a closed, radiation-filled
Friedmann-Robertson-Walker universe, we {\em introduce} and discuss the
supersymmetric double Darboux method in quantum cosmology. A one-parameter family
of ``quantum" universes and the corresponding ``wavefunctions of the universe" for
this case are presented.

\end{abstract}
\bigskip
\bigskip
\begin{center}
{\it based on an essay written for the 1996 Awards of the Gravity Research
 Foundation

(Wellesley Hills, Ma, 02181-0004)}
\end{center}

\bigskip
{\bf Phys. Lett. A 223, 28-30 (25 Nov 1996)}

PACS number(s):  98.80.Hw; 11.30.Pb

\vskip 1cm

%%%%%%%%%%%%%%%%%%%%%%%%        THE ESSAY        %%%%%%%%%%%%%%%%%%%%%%%%
%%%%%%%%%%%%%%%%%%%%%%%%%  written by H.C. Rosu  %%%%%%%%%%%%%%%%%%%%%%%%%%%
%%%%%%%%%%%%%%%%%%%%%%%%%%%% February of 1996 %%%%%%%%%%%%%%%%%%%%%%%%%
%%%%%%%%%%%%%%%%%%%%%%%%%%%% January of 1997 %%%%%%%%%%%%%%%%%%%%%%%%%%%%%%%  

%\section{Introduction}

Many interesting results have been obtained mostly in one-dimensional
quantum mechanics by means
of Darboux-Witten (DW) supersymmetric procedures; for a recent review
see \cite{susy}. These are, essentially,
factorizations of the one-dimensional Schr\"odinger differential operator,
first performed in the supersymmetric context by Witten in 1981 \cite{w81}, 
and in mathematics literature, in the broader sense
of the Darboux transformation,
by Darboux, as early as 1882 \cite{d82}. Constructing families of
isospectral potentials is an important supersymmetric
topic which may have many physical applications \cite{susy}.
Since in quantum cosmology the Wheeler-DeWitt (WDW) equation is a
Schr\"odinger-type
equation one might think of introducing the isospectral issue in this field
as well. This is our purpose in the present work. To accomplish it,
we shall use the simple example of a radiation-filled, closed
Friedmann-Robertson-Walker (FRW) ``quantum" universe.

%\section{Supersymmetric strictly-isospectral potentials on the half-line}

We briefly review the well-established supersymmetric double DW
technique of deleting followed
by ``reinstating" a nodeless, radial (half-line) ground-state, say
$u_0(\tilde{a})$, of a potential $V^{-}(\tilde{a})$ by
means of which one can generate a one-parameter family of isospectral
potentials
$V_{iso}(\tilde{a};\lambda)$, where $\lambda$ is a labeling,
real parameter of each member potential in the set.
The family reads explicitly
%%%%%%%%%%%%%%%%%%
$$
V_{iso}(\tilde{a};\lambda)=V^{-}(\tilde{a})-2[\ln({\cal J} _0+\lambda)]''=
V^{-}(\tilde{a})-\frac{4u_0u_0^{'}}{{\cal J} _0+\lambda}+
\frac{2u_0^4}{({\cal J} _0+\lambda)^2},
\eqno(1)
$$
where the primes are derivatives with respect to the radial
variable $\tilde{a}$ and
%%%%%%%%%%%%%%%%%%
$$
{\cal J} _0(\tilde{a})\equiv\int_0^{\tilde{a}}u_0^2(y)dy.
\eqno(2)
$$
%%%%%%%%%%%%%%%%%%

The result of Eq.~(1) is obtained if one factorizes the
one-dimensional
Schr\"odinger equation with the operators $A=\frac{d}{d\tilde{a}}+
{\cal W}(\tilde{a})$ and
$A^{\dagger}=-\frac{d}{d\tilde{a}}+{\cal W}(\tilde{a})$,
where the superpotential function
is given by ${\cal W}=-u_0^{'}/u_0$. Then, in the DW scheme the potential
and superpotential enter an initial `bosonic' Riccati equation
$V^{-}={\cal W} ^2-{\cal W} ^{'}$. At the same time, one can build a `fermionic'
Riccati equation $V^{+}={\cal W} ^2+{\cal W} ^{'}$, corresponding to a
`fermionic' Schr\"odinger equation for which the operators $A$ and $A^{+}$
are applied in reversed order. Thus, the `fermionic' potential
is found to be $V^{+}=V^{-}(\tilde{a})-2\left (\frac{u_0^{'}}{u_0}
\right)^{'}$.
This potential does not have $u_0$ as the ground state eigenfunction.
However, it is possible
to ``reintroduce" the $u_0$ solution into the spectrum by means of
the general superpotential solution of the fermionic Riccati
equation. The general Riccati solution reads
%%%%%%%%%%%%%%%
$$
{\cal W} _{gen}={\cal W}(\tilde{a}) + \frac{d}{d\tilde{a}}
\ln [{\cal J} _0(\tilde{a})+\lambda]~.
\eqno(3)
$$
%%%%%%%%%%%%%%%%
The way to obtain Eq.~(3) is simple and well-known \cite{mn}.
From
Eq.~(3) one can easily get Eq.~(1) by just using the bosonic Riccati
equation. But, is there any price for this procedure ?
Yes, there is. The new wavefunction differs from the initial one
by a `normalization', $\lambda$-dependent constant. The important thing which was
noticed in the literature was the damping nature of the integral
${\cal J} _0$, both for the
family of potentials as for the wavefunctions. The parameter $\lambda$ is
acting as a ``damping distance"
which is signaling the importance of the integral term. Finally another
useful formula gives the family of
$\lambda$-dependent (non-normalizable) wavefunctions as follows
%%%%%%%%%%%%%%%%%
$$
u_{gen}(\tilde{a};\lambda)=\frac{u_0(\tilde{a})}{{\cal J} _{0} +\lambda}.
\eqno(4)
$$
%%%%%%%%%%%%%%%%%

%\section{WDW equation for the closed, radiation-filled FRW model}

We choose now one of the most simple cosmological metrics, a closed FRW model
filled only with radiation, in order to clearly illustrate the supersymmetric
isospectral issue in quantum cosmology.
 Kung \cite{K} has recently discussed the WDW equation for the closed,
$k >0$ FRW metric with various combinations of
cosmological constant and matter.
The WDW equation used by Kung reads

%%%%%%%%%%%%%%%%%%
$$
\Bigg[-a^{-p}\frac{\partial}{\partial a}a^p\frac{\partial}{\partial a} +
\left(\frac{3\pi}{2G}\right)^2\frac{1}{k^3}\left(ka^2-\frac{8\pi G}{3}
\rho a^4\right)
\Bigg]u(a)=0~,
\eqno(5)
$$
%%%%%%%%%%%%%%%%%%
where the variable $a$ is the cosmological scale factor.
The parameter $p$ enters as a consequence of the ambiguity in the ordering
of $a$ and $\partial/\partial a$; $p=1$ is the so-called Laplacian
ordering, whereas $p=2$ is convenient for the WKB approximation.

In the matter sector various contributions may be present but we shall consider
only the radiation case, i.e.,

%%%%%%%%%%%%%%%%%%
$$
\frac{8\pi G}{3}\rho\rightarrow \frac{8\pi G}{3}\Bigg[\rho _r
\left(\frac{a_0}{a}\right)^4\Bigg]~,
%+ \rho _d\left(\frac{a_0}{a}\right)^3\Bigg]+
%\frac{\Lambda}{3}~,
\eqno(6)
$$
%%%%%%%%%%%%%%%%%%
where $\rho _r$ is the radiation energy density.
The most compact form of the WDW equation in the FRW metric can be written down
as follows,

%%%%%%%%%%%%%%%%%%%%%
$$
\Bigg[-\tilde{a}^{-p}\frac{\partial}{\partial \tilde{a}}\tilde{a} ^p
\frac{\partial}{\partial \tilde{a}} + V(\tilde{a}) \Bigg] u (\tilde{a})=0~,
\eqno(7)
$$
%%%%%%%%%%%%%%%%%%%%%
where the tilde variables are rescaled and dimensionless ones.
Thus, only with the radiation matter sector
the cosmological potential turns out to be an oscillator potential
of the form

%%%%%%%%%%%%%%%%%%%
$$
V(\tilde{a})\equiv\tilde{a} ^2   -
\tilde{\beta}^2~,
\eqno(8)
$$
%%%%%%%%%%%%%%%%%%%
where $\tilde{a} ^2=\frac{3\pi}{2Gk}a^2$ is the tilde scale factor of the
universe, and $\tilde{\beta} ^2=\frac{4\pi ^2}{k^2}\rho _ra_0^4$ expresses
the radiation effect on the cosmological expansion.

With the ansatz $u(\tilde{a})\equiv g(\tilde{a})e^{-\tilde{a}^2/2}$, and
with the further change of variable $x=\tilde{a}^2$, one gets the following 
confluent hypergeometric equation for $g(x)$,

%%%%%%%%%%%%%%%%%%
$$
x\frac{d^2g}{dx^2}+\left(\frac{1+p}{2}-x\right)\frac{dg}{dx}+
\left(\frac{\tilde{\beta}^2-1-p}{4}\right)g=0~.
\eqno(9)
$$
%%%%%%%%%%%%%%%%%%%

Using the $p=1$ factor ordering and with the natural assumption that the
physical
$u$ must vanish asymptotically for $x\rightarrow \infty$, Eq.~(9) turns
into an eigenvalue problem for the Laguerre polynomials
$g(x)=L_{n=(\tilde{\beta}^2-2)/4}^{0}(x)$, where $n\in N_{+}$.
Thus, in this case, the wavefunctions of the universe are of the type

%%%%%%%%%%%%%%%%%%
$$
u_n(\tilde{a})=e^{-\tilde{a}^2/2}L_{(\tilde{\beta}^2-2)/4}^{0}(\tilde{a}^2)~.
\eqno(10)
$$
%%%%%%%%%%%%%%%%%%%
One can also see that the radiation energy density is quantized through the
condition $\tilde{\beta} ^2=4n+2$. Thus,
$\tilde{\beta} =\sqrt{2}$ is the lowest value,
corresponding to the Gaussian ground state wavefunction
$u_0(\tilde{a})=e^{-\tilde{a}^2/2}$. Other factor orderings lead to various confluent
hypergeometric functions $g$ and to a change of the quantization condition
\cite{K}.

%\section{Isospectral, $k>0$, $\rho=\rho _{rad}$ FRW models}

In the following, we shall perform the double Darboux construction using the
Gaussian nodeless wavefunction of the closed, radiation-filled FRW universe.
One can easily obtain ${\cal W}=\tilde{a}$, that is the superpotential is
the dimensionless scale factor of the universe,
and $V^{+}=V^{-}+2=\tilde{a}^2- \tilde{\beta}^2+2=\tilde{a}^2$.

In this case, the one-parameter family of bosonic cosmological potentials can be 
written as follows,
%%%%%%%%%%%%%%%%%%%%%%%%%%%%%%%%%%%%%%%%%%%%%
$$
V_{iso}=\tilde{a}^2-2+
\frac{4\tilde{a}u_0^2}{\lambda _s {\rm erf}(\tilde{a})+\lambda}
+\frac{2u_0^4}{(\lambda _s {\rm erf}(\tilde{a})+\lambda)^2}~.
\eqno(11)
$$
%%%%%%%%%%%%%%%%%%%%%%%%%%%%%%%%%%%%%%%%%%%%
In order to avoid singularities one should take $|\lambda|>\lambda _{s}
\equiv\sqrt{\pi}/2$
\cite{mn}. We have plotted some members of the family equation (11) in Fig.~1.
For $\lambda\rightarrow\pm\infty$ the members are very close in shape to
the original $V^{-}$ potential.

The one-parameter family of wavefunctions of the universe are
%%%%%%%%%%%%%%%%%%%%%%%%%%%%%%%%%%%%
$$
u_{gen}(\tilde{a};\lambda)=\frac{u_0}{\lambda _s {\rm erf}(\tilde{a})+
\lambda}~.
\eqno(12)
$$
%%%%%%%%%%%%%%%%%%%%%%%%%%%%%%%%%%%%%%%%%
The members of the family of wavefunctions given by Eq.~(12) are plotted in Fig.~2
for the same $\lambda$ values as for the potentials. They can also be written as
``Gaussians"
$u_{gen}=e^{-\tilde{a}^2/2\sigma ^2(\tilde{a})}$
with a dispersion depending on the scale factor of the universe
%%%%%%%%%%%%%%%%%%%%%%%%%%%%%%%%%%%%%%%%
$$
\sigma ^2(\tilde{a})=\frac{\tilde{a} ^2}{\tilde{a} ^2+
\ln(\lambda _s {\rm erf}(\tilde{a})+\lambda)^2}~.
\eqno(13)
$$
%%%%%%%%%%%%%%%%%%%%%%%%%%%%%%%%%%%%%%%%%%%%%
We recall that Kiefer \cite{ki} has considered wavepackets with time-dependent 
dispersion within the WKB approximation to minisuperspace models.

In conclusion, we have introduced concepts like isospectral cosmological
potentials and a corresponding one-parameter family of wavefunctions of the
universe for
the simple case of closed, radiation-filled FRW ``quantum" universe.

The work must be considered as merely a brief introduction of the isospectral
supersymmetric problem in quantum cosmology. Its further applications in
this field will be fully explored in future investigations.

%%%%%%%%%%%%%%%%%%%%%%%%%%%%%%%%%%%%%%%%%%%%%%%%%%%%%%%%%%%%%%%%%%%%%
%\section*{Acknowledgment}

\bigskip
\bigskip

This work was partially supported by the CONACyT Project 4868-E9406.

%%%%%%%%%%%%%%%%%%%%%%%%%%%%%%%%%%%%%%%%%%%%%%%%%%%%%%%%%%%%%%%%%%%%%%

%%%%%%%%%%%%%%%%%%%%%%%%%%%%%%%%%%%%%%%%%%%%%%%%%%%%%%%%%%%%%%%%%%%%%%%%%
%\begin{figure}
%\psfig
%{figure=nefami.ps,height=1.8in,bbllx=100bp,bblly=205bp,bburx=360bp,bbury=385bp}

%{\bf Figure Caption}
%\bigskip
%\bigskip

%Fig.~1  Members of the one-parameter isospectral family of
%closed, radiation-filled FRW potentials for the following values of the
%$\lambda$-parameter (from the right to the left):
% (a) -.8863; (b) -.8903; (c) -.9303 together with the
%original potential (the evolvent) $V_{-}=\tilde{a}^2-2$.
%\end{figure}
%\end{document}
%%%%%%%%%%%%%%%%%%%%%%%%%%%%%%%%%%%%%%%%%%%%%%%%%%%%%%%%%%%%%%%%%%%%%%%%%%%%%%%%%%%%%%%%%%%%%%%%%%%%%%%%%%%%%%%%%%%%%%%%%%%%%%%%%%%%%%%%%%%%%%%%%%%%%%%%%%%%%%%%%%%%%%%%%%%%%%%%%%%%%%%%%%%%%%%%%%%%%%%%%%%%%%%%%%%%%%%%%%%%%%%%%%%%%%%%%%%%%%%%%%%%%%%%%%%%%%%%%%%%%%%%%%%%%%%%%%%%%%%%%%%%%%%%%%%%%%%%%%%%%%%%%%%%%%%%%%%%%%%%%%%%%%%%%%%%%%%%%%%%%%%%%%%%%%%%%%%%%%%%%%%%%%%%%%%%%%%%%%%%%%%%%%%%%%%%%%%%%%%%%%%%%%%%%%%%%%%%%%%%%%%%%%%%%%%%%%%%%%%%%%%%%%%%%%%%%%%%%%%%%%%%%%%%%%%%%%%%%%%%%%%%%%%%%%%%%%%%%%%%%%%%%%%%%%%%%%%%%%%%%%%%%%%%%%%%%%%%%%%%%%%%%%%%%%%%%%%%%%%%%%%%%%%%%%%%%%%%%%%%%%%%%%%%%%%%%%%%%%%%%%%%%%%%%%%%%%%%%%%%%%%%%%%%%%%%%%%%%%%%%%%%%%%%%%%%%%%%%%%%%%%%%%%%%%%%%%%%%%%%%%%%%%%%%%%%%%%%%%%%%%%%%%%%%%%%%%%%%%%%%%%%%%%%%%%%%%%%%%%%%%%%%%%%%%%%%%%%%%%%%%%%%%%%%%%%%%%%%%%%%%%%%%%%%%%%%%%%%%%%%%%%%%%%%%%%%%%%%%%%%%%%%%%%%%%%%%%%%%%%%%%%%%%%%%%%%%%%%%%%%%%%%%%%%%%%%%%%%%%%%%%%%%%%%%%%%%%%%%%%%%%%%%%%%%%%%%%%%%%%%%%%%%%%%%%%%%\newpage
\newpage
\centerline{
\epsfxsize=280pt
\epsfbox{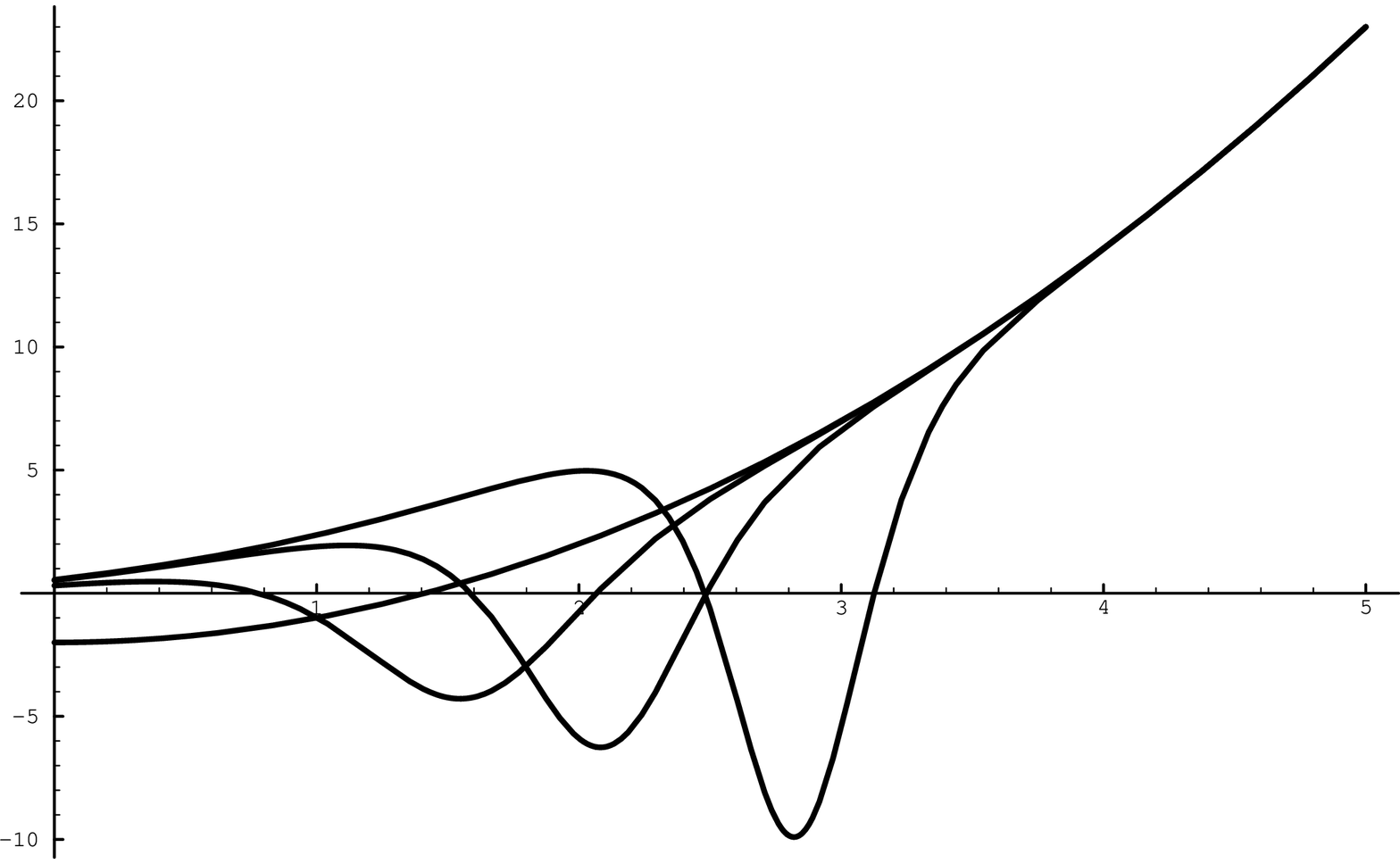}}
\vskip 4ex
\begin{center}
{\small{Fig. 1}\\
 Members of the one-parameter isospectral family of
closed, radiation-filled FRW potentials for the following values of the
$\lambda$-parameter (from the right to the left):
 (a) -.8863; (b) -.8903; (c) -.9303 together with the
original potential (the evolvent) $V^{-}{-}=\tilde{a}^2-2$.}
\end{center}

\vskip 2ex
\centerline{
\epsfxsize=280pt
\epsfbox{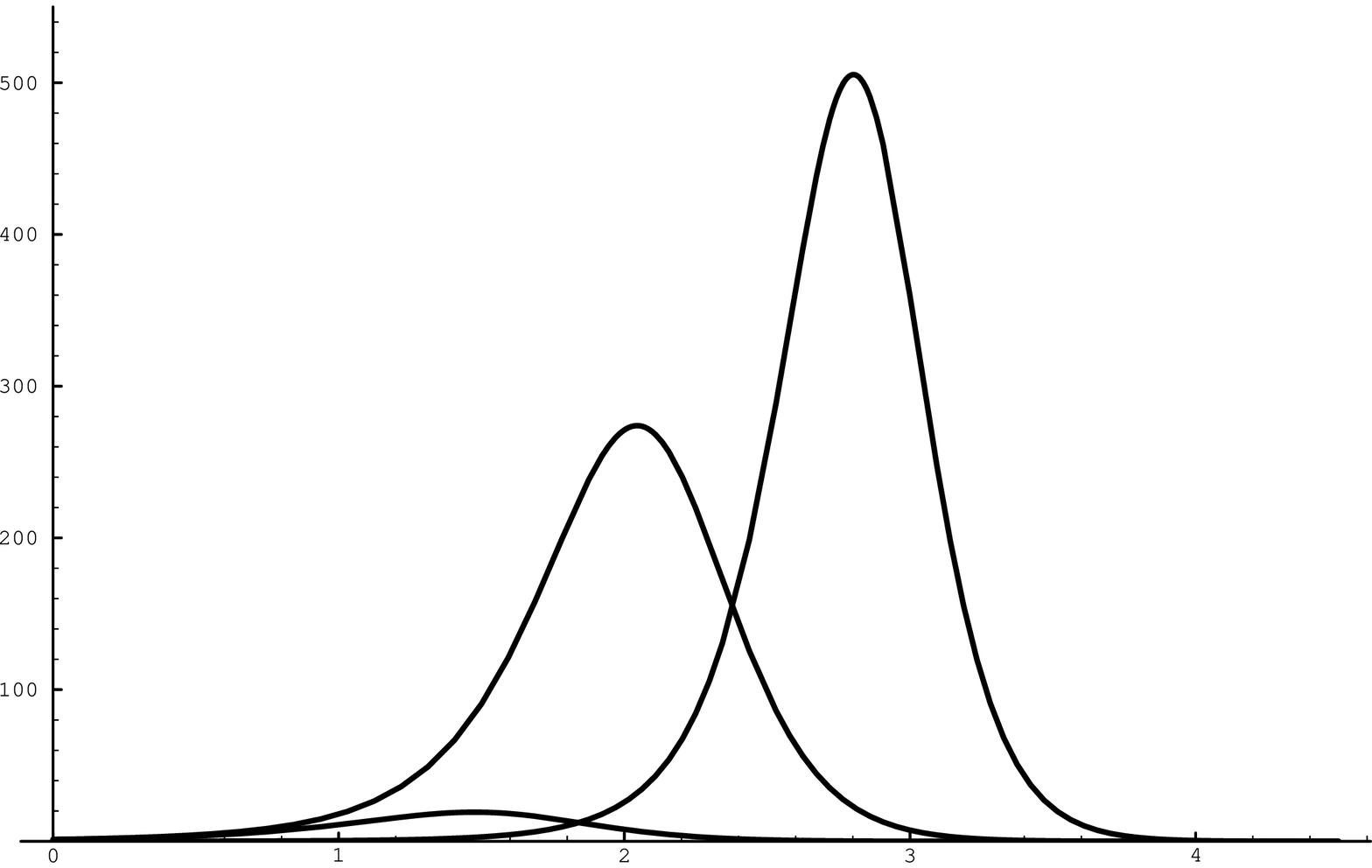}}
\vskip 4ex
\begin{center}
{\small{Fig. 2}\\
Squares of the ``wavefunction of the universe" for the same values (right to
left) of the $\lambda$ 
parameter as in Fig.~1. The largest peak is scaled down forty times.}
\end{center}

\end{document}